\documentclass[12pt]{article}
\usepackage{epsfig}
\usepackage{graphicx}
\usepackage{amsmath}
\usepackage{hhline}
\usepackage{amssymb}
\usepackage{times}
\usepackage{cite}

\newlength{\dinwidth}
\newlength{\dinmargin}
\begin{document}
\newcommand{\pom}{{I\!\!P}}
\newcommand{\reg}{{I\!\!R}}
\newcommand{\slowpi}{\pi_{\mathit{slow}}}
\newcommand{\fiidiii}{F_2^{D(3)}}
\newcommand{\fiidiiiarg}{\fiidiii\,(\beta,\,Q^2,\,x)}
\newcommand{\n}{1.19\pm 0.06 (stat.) \pm0.07 (syst.)}
\newcommand{\nz}{1.30\pm 0.08 (stat.)^{+0.08}_{-0.14} (syst.)}
\newcommand{\fiidiiiful}{F_2^{D(4)}\,(\beta,\,Q^2,\,x,\,t)}
\newcommand{\fiipom}{\tilde F_2^D}
\newcommand{\ALPHA}{1.10\pm0.03 (stat.) \pm0.04 (syst.)}
\newcommand{\ALPHAZ}{1.15\pm0.04 (stat.)^{+0.04}_{-0.07} (syst.)}
\newcommand{\fiipomarg}{\fiipom\,(\beta,\,Q^2)}
\newcommand{\pomflux}{f_{\pom / p}}
\newcommand{\nxpom}{1.19\pm 0.06 (stat.) \pm0.07 (syst.)}
\newcommand {\gapprox}
   {\raisebox{-0.7ex}{$\stackrel {\textstyle>}{\sim}$}}
\newcommand {\lapprox}
   {\raisebox{-0.7ex}{$\stackrel {\textstyle<}{\sim}$}}
\def\gsim{\,\lower.25ex\hbox{$\scriptstyle\sim$}\kern-1.30ex%
\raise 0.55ex\hbox{$\scriptstyle >$}\,}
\def\lsim{\,\lower.25ex\hbox{$\scriptstyle\sim$}\kern-1.30ex%
\raise 0.55ex\hbox{$\scriptstyle <$}\,}
\newcommand{\pomfluxarg}{f_{\pom / p}\,(x_\pom)}
\newcommand{\dsf}{\mbox{$F_2^{D(3)}$}}
\newcommand{\dsfva}{\mbox{$F_2^{D(3)}(\beta,Q^2,x_{I\!\!P})$}}
\newcommand{\dsfvb}{\mbox{$F_2^{D(3)}(\beta,Q^2,x)$}}
\newcommand{\dsfpom}{$F_2^{I\!\!P}$}
\newcommand{\gap}{\stackrel{>}{\sim}}
\newcommand{\lap}{\stackrel{<}{\sim}}
\newcommand{\fem}{$F_2^{em}$}
\newcommand{\tsnmp}{$\tilde{\sigma}_{NC}(e^{\mp})$}
\newcommand{\tsnm}{$\tilde{\sigma}_{NC}(e^-)$}
\newcommand{\tsnp}{$\tilde{\sigma}_{NC}(e^+)$}
\newcommand{\st}{$\star$}
\newcommand{\sst}{$\star \star$}
\newcommand{\ssst}{$\star \star \star$}
\newcommand{\sssst}{$\star \star \star \star$}
\newcommand{\tw}{\theta_W}
\newcommand{\sw}{\sin{\theta_W}}
\newcommand{\cw}{\cos{\theta_W}}
\newcommand{\sww}{\sin^2{\theta_W}}
\newcommand{\cww}{\cos^2{\theta_W}}
\newcommand{\trm}{m_{\perp}}
\newcommand{\trp}{p_{\perp}}
\newcommand{\trmm}{m_{\perp}^2}
\newcommand{\trpp}{p_{\perp}^2}
\newcommand{\alp}{\alpha_s}

\newcommand{\alps}{\alpha_s}
\newcommand{\sqrts}{$\sqrt{s}$}
\newcommand{\LO}{$O(\alpha_s^0)$}
\newcommand{\Oa}{$O(\alpha_s)$}
\newcommand{\Oaa}{$O(\alpha_s^2)$}
\newcommand{\PT}{p_{\perp}}
\newcommand{\JPSI}{J/\psi}
\newcommand{\sh}{\hat{s}}
\newcommand{\uh}{\hat{u}}
\newcommand{\MP}{m_{J/\psi}}
\newcommand{\PO}{I\!\!P}
\newcommand{\xbj}{x}
\newcommand{\xpom}{x_{\PO}}
\newcommand{\ttbs}{\char'134}
\newcommand{\xpomlo}{3\times10^{-4}}
\newcommand{\xpomup}{0.05}
\newcommand{\dgr}{^\circ}
\newcommand{\pbarnt}{\,\mbox{{\rm pb$^{-1}$}}}
\newcommand{\gev}{\,\mbox{GeV}}
\newcommand{\WBoson}{\mbox{$W$}}
\newcommand{\fbarn}{\,\mbox{{\rm fb}}}
\newcommand{\fbarnt}{\,\mbox{{\rm fb$^{-1}$}}}
%
%
\newcommand{\qsq}{\ensuremath{Q^2} }
\newcommand{\gevsq}{\ensuremath{\mathrm{GeV}^2} }
\newcommand{\et}{\ensuremath{E_t^*} }
\newcommand{\rap}{\ensuremath{\eta^*} }
\newcommand{\gp}{\ensuremath{\gamma^*}p }
\newcommand{\dsiget}{\ensuremath{{\rm d}\sigma_{ep}/{\rm d}E_t^*} }
\newcommand{\dsigrap}{\ensuremath{{\rm d}\sigma_{ep}/{\rm d}\eta^*} }
\newcommand{\dedx}{\ensuremath{{\rm d} E/{\rm d} x}}
\def\Journal#1#2#3#4{{#1} {\bf #2} (#3) #4}
\def\NCA{Nuovo Cimento}
\def\RPP{Rep. Prog. Phys.}
\def\ARNPS{Ann. Rev. Nucl. Part. Sci.}
\def\NIM{Nucl. Instrum. Methods}
\def\NIMA{{Nucl. Instrum. Methods} {\bf A}}
\def\NPB{{Nucl. Phys.}   {\bf B}}
\def\NPPS{Nucl. Phys. Proc. Suppl.}
\def\NPPSC{{Nucl. Phys. Proc. Suppl.} {\bf C}}
\def\PR{Phys. Rev.}
\def\PLB{{Phys. Lett.}   {\bf B}}
\def\PRL{Phys. Rev. Lett.}
\def\PRD{{Phys. Rev.}    {\bf D}}
\def\PRC{{Phys. Rev.}    {\bf C}}
\def\ZPC{{Z. Phys.}      {\bf C}}
\def\EJC{{Eur. Phys. J.} {\bf C}}
\def\EPL{{Eur. Phys. Lett.} {\bf}}
\def\CPC{Comp. Phys. Commun.}
\def\NP{{Nucl. Phys.}}
\def\JPG{{J. Phys.} {\bf G}}
\def\EPC{{Eur. Phys. J.} {\bf C}}
\def\PRSL{{Proc. Roy. Soc.}} {\bf}
\def\PETF{{Pi'sma. Eksp. Teor. Fiz.}} {\bf}
\def\JETPL{{JETP Lett}}{\bf}
\def\IJTP{Int. J. Theor. Phys.}
\def\HJ{Hadronic J.}




\begin{titlepage}

\vspace*{2.cm}
\begin{center}
\begin{Large}
{\boldmath \bf Testing the proposed causal link between cosmic rays and cloud cover} \\

\end{Large}

\vspace*{1.2cm}

T. Sloan $^{1}$, A.W.Wolfendale$^{2}$ \\

\vspace{1cm}
\ $^1$Physics Department, University of Lancaster, Lancaster, UK\\
\ $^2$Physics Department, Durham University, Durham, UK\\
\ email t.sloan@lancaster.ac.uk \\

\end{center}

\vspace*{.5cm}

\begin{abstract}
A decrease in the globally averaged low level cloud cover, deduced 
from the ISCCP infra red data, as the cosmic ray intensity decreased 
during the solar cycle 22 was observed by two groups. The groups 
went on to hypothesise that the decrease 
in ionization due to cosmic rays causes the decrease in cloud cover, 
thereby explaining a large part of the presently observed global 
warming. We have examined this hypothesis to look for evidence to 
corroborate it. None has been found and so our conclusions are to 
doubt it. From the absence of corroborative evidence, we estimate 
that less than $23\%$, at the $95\%$ confidence level, of the 
11-year cycle changes in the globally averaged cloud cover observed 
in solar cycle 22 is due to the change in the rate of ionization 
from the solar modulation of cosmic rays. 
\noindent
\end{abstract}


\end{titlepage}






\section{Introduction}

In references \cite{PBB} and \cite{MS1} a correlation was demonstrated 
for `low clouds' ($<$3.2 km in altitude) between the changes in the 
globally averaged  
'low cloud cover (LCC) anomaly' and the changes in the cosmic ray (CR) 
count rate (e.g. see figure 1 of \cite{MS1}). Here 'LCC anomaly' means 
the difference between the
mean monthly LCC and the time averaged value for the month. The LCC
anomaly was derived by these groups from the satellite data 
provided by the  
International Satellite Cloud Climatology Project (ISCCP) from 
the monthly averaged D2 analysis using
the infrared data\cite{ISCCP}. It was implied by both groups that a
decrease in CR intensity causes a decrease in LCC. Since this may
not be the case if both effects are correlated to a third
variable, it is prudent to look for further evidence of such a
causal connection. Such a causal connection would have vast
importance since, according to \cite{PBB,MS1,MS3}, it could be the main 
cause of the presently observed global warming.  
The proposed mechanism for this depends on the observation of an 
increase in solar activity over the last century \cite{Lockwood}. 
An increase in solar activity causes a net decrease in CR intensity 
which, according to the causal connection proposed 
in \cite{PBB,MS1,MS3}, causes a decrease in LCC. This, in turn, leads to
increased warming of the Earth's surface by the Sun. The effects of 
solar changes  
on the increases in the global mean surface air temperature have been 
discussed more fully in \cite{LandF} and are reviewed in \cite{Haigh}. 

The International Panel on Climate Change (IPCC) has not considered 
this effect as significant \cite{IPCC} since the origin of the 
correlation observed 
in \cite{PBB,MS1} has been questioned \cite{KSKK}. The grounds for this 
doubt are that the ISCCP infrared data give different results from the 
day time low cloud data and also 
that the correlation after 1994 is of poor quality. The correlation was 
also questioned since a similar one was observed over the USA but with 
the opposite sign \cite{UC} to that seen in \cite{PBB,MS1}. These doubts 
should be weighed against the following. Firstly, in the daytime LCC 
shown in figure 1b of \cite{KSKK} there is structure at the maximum of 
the solar activity, albeit with a poorer correlation than the infra red 
data with the CR modulation. Secondly, there is no inconsistency between 
the surface data over the USA seen in \cite{UC} and the ISCCP infra red 
data since the latter also show an anti-correlation over the USA (see fig 2a 
in \cite{MS1}, which we also confirm). Thirdly, a correlation between the CR 
rate and cloud cover was also observed in \cite{Harrison} where cloud cover was 
determined in a completely different way from that adopted in \cite{PBB,MS1,MS3}.
Fourthly, a latitude dependence between the calculated ion concentration from CR 
at altitude 3 km and the low cloud amount was reported in selected local regions 
where the correlation coefficient between the two distributions is high \cite{uso}. 
Fifthly, whilst 
after 1994 there is a poor correlation at high Earth latitudes, we see a possible 
correlation in the tropical regions (see below) and it is well known that sequential 
solar cycles behave differently from each other due to the reversal of the solar 
magnetic field. The IPCC labels the level of scientific understanding 
of the observed correlation as ``very low''. Given these facts the correlation observed 
in \cite{PBB,MS1} needs to be studied further. Here we adopt the approach of looking 
for other possible manifestations of the causal connection, assuming that it exists, 
in order to corroborate the effect or otherwise.

The implication of the causal connection proposed in \cite{PBB,MS1} is
that LCC is influenced by the rate of ion production in the atmosphere. 
In this paper, we have examined various incidences of ionizing
radiation changes in the atmosphere from cosmic rays 
to look for consequential changes in
LCC which would result if the causal connection existed. 
We have looked for changes in LCC from changes in 
the CR intensity due to solar activity as
the geomagnetic latitude increases i.e. as the vertical rigidity
cut-off (VRCO) decreases. We have also looked at the effects on 
LCC of the known sporadic changes in the CR intensity.  
These cases, where there is a change in the 
ionization rate, have been examined to see if a corresponding change in  
cloud cover occurs, as would be expected from the 
causal connection hypothesized in \cite{PBB,MS1}. Throughout we use 
the same ISCCP D2 data sample as in \cite{PBB,MS1} unless otherwise stated.  

\section{The Correlation between Cosmic Rays (CR) and Low Cloud Cover (LCC)}
\label{ions}

\begin{figure}[htb]
\vspace*{-25mm}
\begin{center}
\includegraphics [width=0.85\textwidth]{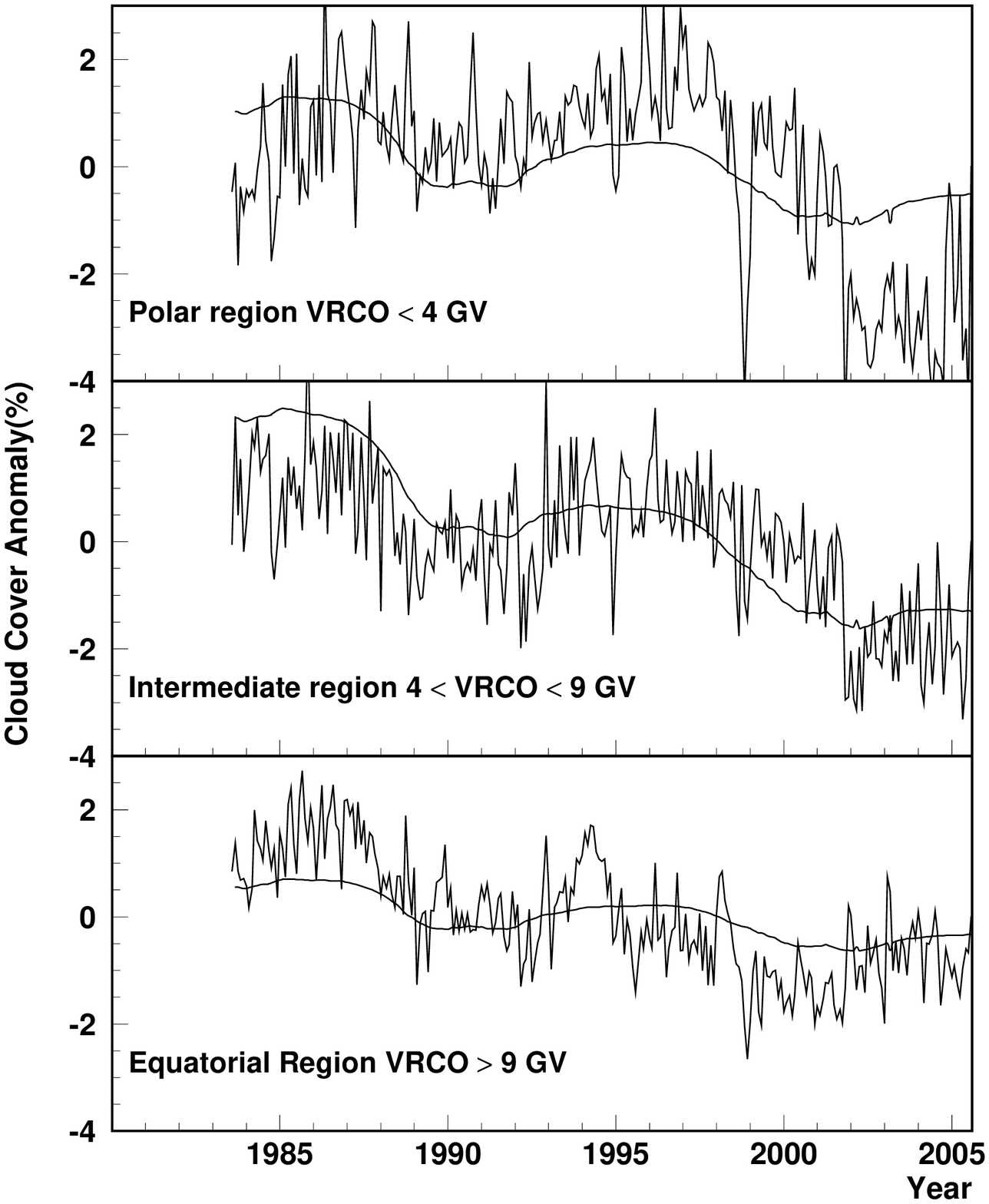}
\end{center}
\vspace*{-8mm}
\caption{The LCC anomaly as a function of time for various
ranges of vertical cut off rigidity (VRCO). The smooth curve 
shows a fit of the monthly mean of the daily sun spot number (SSN) 
with an assumed linearly falling systematic change. The SSN is 
anti-correlated with the CR count rate with a lead time of some  
months}
\label{fig1}
\end{figure}

Figure \ref{fig1} shows the LCC anomaly determined from the ISCCP infra red 
data as a function of time averaged over the Earth, in three separate regions. 
The smooth curves in figure \ref{fig1} show the best fits of the LCC anomaly 
to the mean daily sun spot (SS) number (inverted) superimposed on an assumed 
linear change with time in the LCC anomaly. 
Such a change may be real or it could be due to an artefact  
of the satellite instrumentation as discussed in \cite{evan}. 
The fit was made using the CERN library fitting programme MINUIT
\cite{MINUIT} to minimize the value of $\chi^2$ between the measurements
and the curve. The errors on the data points were taken from the mean
square deviations of independent pairs of neighbouring points.The free 
parameters in the fit were the slope and intercept of the assumed smooth 
linear systematic change in the LCC, a multiplicative constant for 
the monthly averaged daily American sunspot number  (SSN) \cite{USSS}
and a time shift for the delay between the onset of the $dip$ in the LCC and
the increase in the SSN. The multiplicative constant represents the
amplitude of the dip in the LCC per unit change in SSN. We take this
amplitude as the magnitude of `the effect'.  The fits were rather poor
(see figure \ref{fig1}) with values of $\chi^2$ per degree of freedom
of from 1.5 to 2.5. However, fits between 1985 to 1996 (solar cycle 22) 
were better than this, allowing the amplitude of the dip to be determined 
in this time range. The modulation of the cosmic ray intensity 
is strongly anti-correlated with the variation in the SSN. 
The time shift between the onset of the dip and the change in the SSN  
will be used in the manner to be described later.  
 
The observed dip in figure \ref{fig1} is similar to that
seen in \cite{PBB,MS1} between the years 1985 and 1995. However, 
the dip in LCC seen in solar cycle 22 (peaking in 1990) is not evident 
in solar cycle 23 (peaking in 2000) except, surprisingly, in the equatorial 
region where the solar modulation is least.

The globally averaged decrease in LCC during solar cycle 22 (averaging 
the dips in figure \ref{fig1}) is $1.28\pm0.14\%$. The globally averaged total 
LCC amount is $28\%$ giving a change in LCC during the dip of 
$\Delta \rm{LCC}/\rm{LCC}=4.6\pm0.5\%$. The globally averaged peak to peak 
modulation in the CR neutron monitor count rate is computed to be 
$11\pm1\%$ 
of the total. The neutron modulation was determined 
from a study of the data from 35 neutron monitors around the 
globe \cite{Watanabe} using similar methods to those described 
in \cite{Braun}. A fit to the measurements of the peak to peak 
modulation versus SSN gives 
$\Delta N/N=1.15~10^{-3}-0.061~10^{-3}\cdot V$ per SSN,  
where $V$ is the VRCO
The muon modulation is a factor 3 lower than this \cite{Allkofer}
due to the higher primary energy needed to produce muons. Ionization
is also produced from the electromagnetic component of CR whose
long term modulation has not been measured. This will depend on $\pi^0$
production from CR primary interactions which will have a threshold
energy intermediate between those for muons and neutrons. We therefore
assume that the total globally averaged solar modulation of the cosmic 
ray ionization rate is the average of those for muons and neutrons i.e.
$7\pm3\%$, where the uncertainty bridges the gap from muons to neutrons.  
The solar modulation of the globally averaged ionization rate, $q$, will be 
reduced to $\Delta q/q = 6\pm3\%$ by the dilution of the ionization over 
land ($29\%$ of the Earth's surface) by radioactivity which will produce an
ionization rate of a similar magnitude to that from CR at low cloud altitude 
\cite{Karunakara}. The fractional change in the LCC is therefore 
related to the rate of ionization change  due to solar modulation by   
\begin{equation}
\frac{\Delta \rm{LCC}}{\rm{LCC}} = 0.77\pm0.38\cdot\frac{\Delta q}{q} 
\end{equation} 
implying that $\rm{LCC}\propto q^\xi$ with $\xi=0.77\pm0.38$, where the 
error is dominated by the uncertainty in $\Delta q/q$.  
This is compatible, within the error, with a $q^{0.5}$ behaviour. 
Such behaviour is expected, at least in clean air,   
if $\rm{LCC} \propto n$, where n is the small ion concentration which is 
expected to be limited mainly by recombination \cite{Mason}.  

\begin{figure}[htb]
\vspace*{-25mm}
\begin{center}
\includegraphics [width=1.0\textwidth]{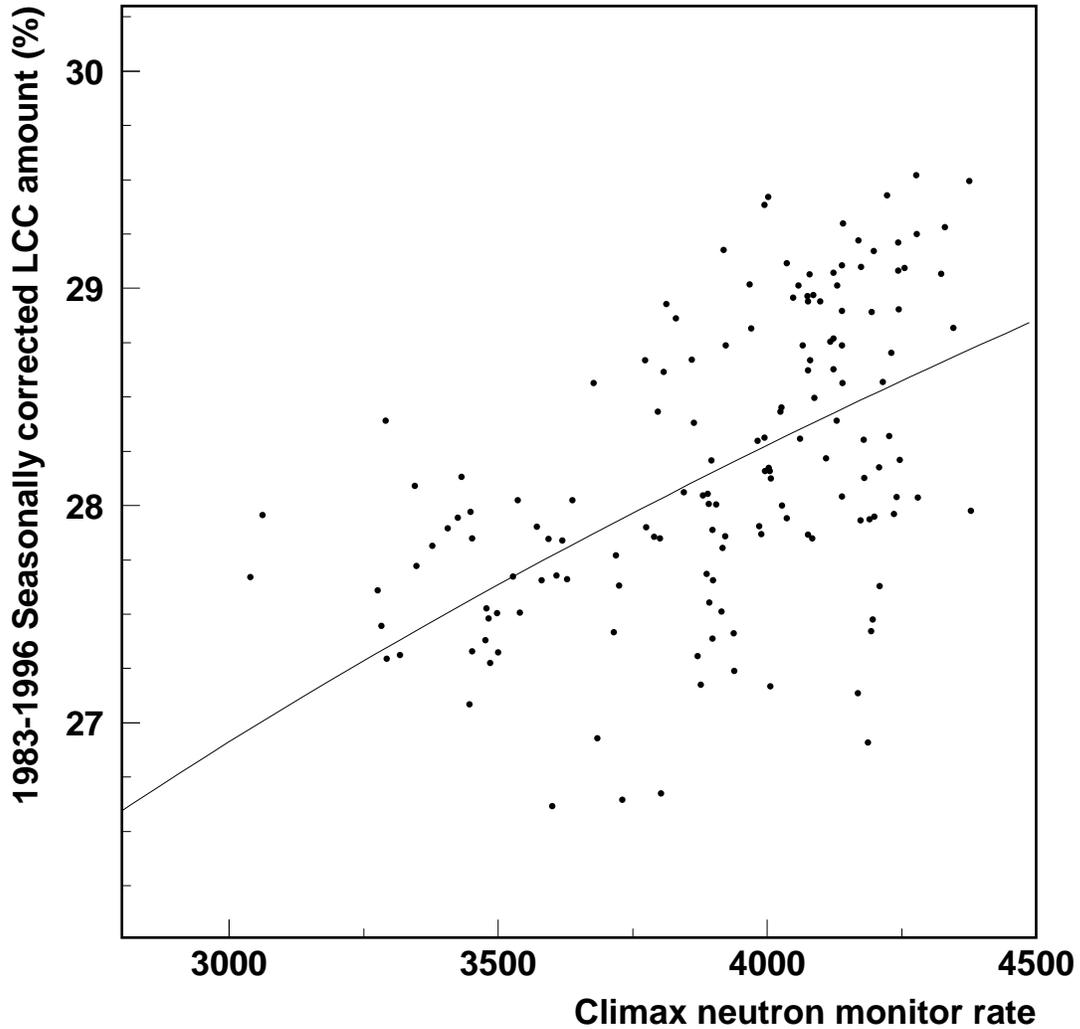}
\end{center}
\vspace*{-8mm}
\caption{The seasonally corrected LCC amount as a function of the Climax neutron 
monitor count rate (both monthly averaged) during solar cycle 
22 (1983-1996). The seasonally corrected LCC amount was obtained by adding the 
globally averaged LCC to the monthly globally averaged anomalies. 
The smooth curve shows the fit described in the text.}
\label{figsec}
\end{figure}

To study the detailed shape of the correlation shown in figure \ref{fig1} 
the globally averaged LCC amount is plotted directly against the Climax neutron 
counter monitor rate, $N_C$, in figure \ref{figsec}. The good correlation is evident. 
Fits of the form  
\begin{equation}
\rm{LCC}=\beta+\gamma N_C^{\alpha}
\label{fit}
\end{equation} 
have been made. Here the first parameter, $\beta$, can be interpreted as a measure of the 
LCC amount attributable to non-ionizing sources and the second term to ionizing sources. 
If the part of the LCC amount which depends on ionization is proportional to the small 
ion concentration, $n$, and $n\propto q^\xi$, the parameter $\alpha$ 
is related to $\xi$ by 
$\alpha=a_1a_2\xi$, where $a_1=(\delta q/q)/(\delta N/N)$ and  
$a_2=(\delta N/N)/(\delta N_C/N_C)$. \footnote{It can be seen that 
$\delta q/q=a_1 a_2 \delta N_C/N_C$ which on integration gives $q\propto N_C^{a_1a_2}$ 
i.e. $q^\xi\propto N_C^{a_1a_2\xi}$.}  
We take $\delta q/q=6\pm3\%$, the globally averaged solar modulation $\delta N/N=11\pm1\%$ 
(as discussed above) and the solar modulation measured for the Climax detector to 
be $\delta N_C/N_C=19\%$, so that $a_1a_2=0.32\pm0.16$.


The data in figure \ref{figsec} are insufficient to determine 
precisely the parameters, $\alpha, \beta$ and $\gamma$ separately. 
Fits with different combinations of the parameters are equally good as measured by  
the $\chi^2$. However, the value of $\chi^2$ rapidly becomes unacceptable when  
$\beta$ is increased to a value corresponding to more than $70\%$ of the cloud 
arising from non-ionizing sources, i.e. a fraction of at least $30\%$ comes 
from ionization. The following argument also shows that 
the latter fraction must be large.  
The smooth curve in figure \ref{figsec} shows the fit with $\beta=0$ 
which gives $\alpha=0.17$ with a $\chi^2=148.9$ for 146 degrees of freedom and 
a correlation coefficient of 0.54. The values of the parameters $\alpha$, $\beta$ 
and $\gamma$ are strongly correlated such that increasing values of $\alpha$ are 
associated with increasing values of $\beta$ and decreasing values of $\gamma$.    
The fits with $\alpha > 0.16$, corresponding to $\xi > 0.5$, give positive values of 
$\beta$ while fits with  $\alpha < 0.16$, corresponding to values of $\xi <0.5$, 
give negative values of $\beta$ which are unphysical. Assuming that it is 
implausible that the LCC amount generated by ionization  varies faster than 
linearly with $q$, i.e. $\xi < 1$, then $\alpha$ must be less than 0.48, 
taking $\delta q/q$ at its upper limit of $9\%$. Such a value of $\alpha$ 
gives a fit with $\beta = 20\%$, implying that the fraction of the LCC 
generated by sources other than ionization is less than 20/28=0.7 i.e. a 
minimum fraction of 0.3 of the LCC amount is generated by ionization. 
At $\xi=0.5$ the value of $\beta$ is compatible with zero. Hence assuming 
$\xi$ lies in the range 0.5 to 1.0 the fraction of the LCC generated by 
ionization lies somewhere between 1 and 0.3, respectively.   

In summary, assuming that the correlation shown in figures \ref{fig1} 
and \ref{figsec} is not accidental 
a very large fraction of the LCC must be generated by ionization. 
We now attempt to corroborate this assumption and this necessary deduction.

\section{Latitude dependence of `the effect'}
\label{lats}
It is well known that the magnitude of the CR time variation,
due to the 11 year solar cycle, varies with latitude.  More
accurately, it is a function of the VRCO, the reason being
that the geomagnetic field deflects away more low energy particles 
as the geomagnetic equator (highest VRCO) is approached.  
Since the CR flux increases rapidly as the primary energy decreases, 
the solar 
modulation becomes less severe as the VRCO increases towards  
the geomagnetic equator. Hence, if the causal connection between 
the CR ionization rate and LCC proposed in \cite{PBB,MS1} exists 
with the necessary large fraction of the LCC produced by ionization 
demonstrated above, 
one would expect larger changes in LCC at low values of VRCO than 
at high values.  Furthermore it is
known that there is a delay of some months between the
decrease in the CR intensity and the increase in the sun spot (SS) 
number with the even numbered solar cycles showing smaller delays
than the odd numbered \cite{Kudela}. Note that the CR count rate
is anti-correlated to the SS number.

\begin{figure}
\begin{center}
\includegraphics [width=1.0\textwidth]{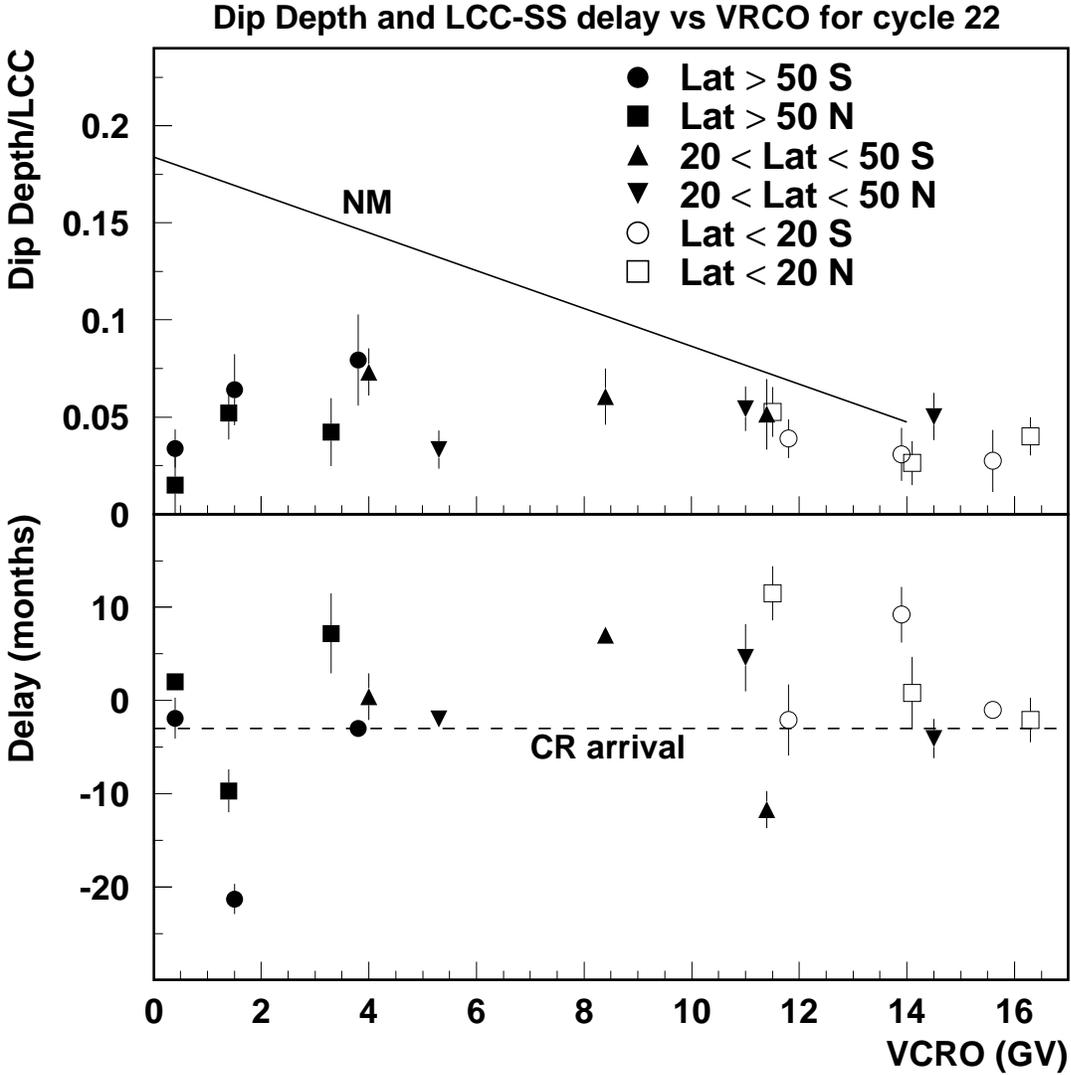}
\end{center}
\vspace*{-8mm}
\caption{ The observed modulation of the LCC (upper panel) as measured
from the fit to solar cycle 22 only (see figure \ref{fig1}). The
`modulations' are expressed by the dip amplitude at the time of the solar 
maximum (1991) divided by the mean LCC. The smooth curve labelled NM
shows a fit to the fractional modulation, $~dN/N$, measured from 
neutron monitors around the World (see text). The lower panel shows
the fitted delay between the onset of the dip and that of the SS
number in months. The dashed line shows the expected delay if a
correlation existed between the changes in CR and CC. The measured
delay between the CR decrease and increase in SSN is 3 months 
in cycle 22. NB positive delay means CC
precedes the increase in SSN.}
\label{fig3}
\end{figure}

The observed dip in figure \ref{fig1} is similar to that
seen in \cite{PBB,MS1} between the years 1985 and 1995. However, the
expected rise in amplitude of this dip with decreasing VRCO is not
apparent. 
To investigate the effect of the VRCO further and to check that the 
above result was not due to a latitude dependent efficiency of the
cloud production mechanism, the LCC was determined in three
strips of latitude for the Northern and Southern hemispheres of the 
Earth separately. The amplitude of the dip in solar cycle 22
was measured from the fit for each, as a function of VRCO. The dip 
was visible in every subdivision. 
Figure \ref{fig3} (upper panel) confirms that the amplitude
of the dip appears to be rather constant with VRCO rather than
increasing with the observed increase in CR modulation determined 
as described above. Furthermore there is no discernible difference 
between the Northern (where oceans are less dominant) and Southern 
hemispheres (where oceans are more dominant).  
Figure \ref{fig3} (lower panel) shows that 
the measured value of the delay between the onset of
the dip and the change in SS number fluctuates randomly rather
than concentrates around a fixed delay (expected to be $-3$ months
for the CR increase in solar cycle 22). Each latitude band has a
median value compatible
with zero with an overall mean of $-0.9\pm1.6$ months,
where the error is the standard error determined from the 
root mean square (RMS) 
deviation of the measurements from the mean. This is
compatible with the onset of the increase in SS number but
somewhat earlier than the arrival time of the CR increase
($-3$ months). Hence there is a somewhat better time correlation
between the start of the dip and onset of the increase in the 
SS number than with the change in the CR rate, although 
the error is too large to be conclusive.  

Neither the amplitude variation with VRCO nor the arrival times
shown in figure \ref{fig3} corroborate the claim of a full causal
connection between CR ionization rate and the LCC anomaly. We proceed 
to set a limit on any contribution from a partial correlation. 

We attempt to quantify the part of the dip related to changes 
in the CR ionization rate and that related to other sources which 
are independent of the ionization rate, as follows. The change in LCC 
during the solar cycle, $\Delta \rm{LCC}$, can be decomposed into a 
part which is dependent on the change in the ionization rate $\Delta \rm{LCC}_I$ 
and a part due to other mechanisms correlated with solar activity 
but not directly due to ionization, $\Delta\rm{LCC}_S$, i.e.  
$\Delta \rm{LCC}=\Delta\rm{LCC}_I+\Delta\rm{LCC}_S$.  
Differentiation shows that $\Delta\rm{LCC}_I=\kappa dN/N$.  
where $\kappa=Ndg(N)/dN$ with    
$g(N)$ the functional dependence of the LCC on the ionization 
rate as measured by the neutron monitor rate, $N$. The function $Ndg(N)/dN$  
is slowly  varying with $N$ for reasonable functions, $g(N)$, over the range of 
changes of $\delta N/N$ during the solar cycle, so that $\kappa$ is approximately 
constant.  For example, if $\rm{LCC} \propto n \propto q^\xi \propto N^{a\xi}$ 
where $a=(\delta q/q)/(\delta N/N) \sim 0.5$ (see above) and $\xi\sim 0.5$,   
$\kappa$ will change by $\sim 5\%$ as $\delta N/N$ changes from 0 to 0.2. 
From this it can be seen that the dip depth may be expressed as   
\begin{equation}
\Delta\rm{LCC}=\Delta\rm{LCC}_S+\kappa \delta N/N
\end{equation}
where $\kappa$ can be treated as a constant. 

We use this to identify the part of the
distribution in the upper panel of figure \ref{fig3} which
correlates with the CR modulation. A fit was performed of the shape
of the neutron modulation variation (the correlated part) and a
constant term (the uncorrelated part) to the measurements. The fit 
gave the fraction of the distribution correlated with the neutron
modulation to be $0.02\pm0.13~$ i.e. compatible with zero
with a value of $\chi^2=17.8$  for 16 degrees of freedom. From
this it is deduced that less than $23\%$ of the distribution, at
the 95$\%$ confidence level, belongs to the part correlated with the
CR modulation and more than $77\%$ belongs to the other sources 
correlated to solar activity but not directly to the change in 
ionization rate.  These limits are incompatible with a large part of the 
change in the LCC during solar cycle 22 being produced by a change in 
ionization and so they do not corroborate the hypothesis of such a 
change proposed in \cite{PBB,MS1}. The correlation seen in figures \ref{fig1} 
and \ref{figsec}, if real, must be due to an effect, other than ionization,  
which is correlated with solar activity. 

This upper limit represents a limit on the fraction of the globally averaged 
dip in the LCC seen in solar cycle 22 which is caused by CR ionization. 
There could be local changes from this ionization such as those reported 
in \cite{uso} which, from the above upper limit, must contribute less than 
a fraction of $23\%$ to the globally averaged dip.    

\section{Sporadic Changes in Cosmic Ray activity}  
\label{searches}


Rapid changes in CR intensity occur from time to time. These take the 
form of large intensity  increases, so called ground level events (GLE), 
or smaller decreases in intensity (Forbush decreases) \cite{Forbush}.  
Such changes of intensity usually last for periods from a few hours to 
days and sometimes longer in the case of Forbush events. A survey has 
been given in \cite{Velinov}. These changes present an opportunity to 
test for LCC - CR correlations since if the causal connection proposed 
in \cite{PBB,MS1} exists one would expect to see changes in the LCC 
at the times of these events. The causal connection implies an increase 
(decrease) in LCC following a GLE (Forbush decrease) and we   
assume that such changes occur in times shorter than days.  

There were 3 very large GLEs during the time span of the ISCCP cloud data (1985-2005), 
each lasting several hours. The event on 29 September 1989 was clearly seen in both 
the CR neutron and muon monitors \cite{duldig}. The peak intensity neutron monitor 
enhancement in this event was observed to change from four times the steady state 
value for neutron monitors with VRCO close to zero down to 1.17 times the steady state 
value at a VRCO of 11.5 GV while the muon monitors varied from 1.4 \cite{duldig} to 
1.08 times the steady state value in the same range of VRCO.  
The other two events (on 24 Oct 1989 and on 20 Jan 2005) had similarly large 
neutron monitor signals but they did not produce visible signals in the Nagoya 
muon monitor\cite{Nagoya}. The global LCC 
averages as a function of time were reconstructed from the ISCCP D1 data, which 
are 3 hour averages rather than the monthly averages of the D2 data, at times before 
and after each of the three GLEs. There were no visible anomalous changes in 
these global averages following each GLE where an increase of more than $2\%$ would 
have appeared anomalous. It is difficult to make quantitative estimates of the expected 
changes in the LCC, according to the hypothesis of \cite{PBB,MS1}, from such events since 
the amount of ionization produced by them is unknown. One can only conclude that the 
events do not provide corroborative evidence for the causal connection between cloud 
cover and ionization proposed in \cite{PBB,MS1,MS3} even though the changes in the 
neutron monitor rate were very large.         

\begin{figure}
\begin{center}
\includegraphics [width=0.9\textwidth]{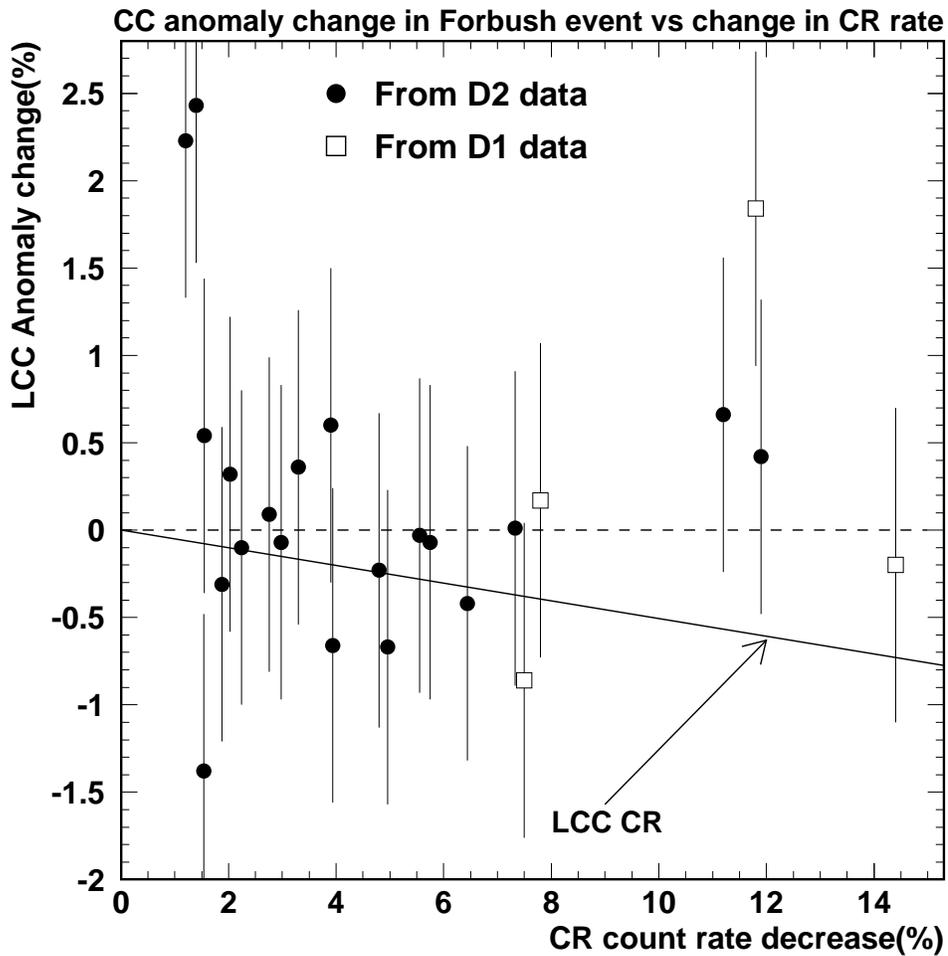}
\end{center}
\vspace*{-8mm}
\caption{The measured change in the LCC plotted against the change in the 
Oulu neutron monitor count rate during the measurement time of 1 month for 
the D2 data (solid circles) and 1 week for the D1 data (open squares).  
The solid line shows the values expected from the smooth curve shown in figure
\ref{figsec}. The Oulu count rate was observed to change by $17\%$ 
due to the solar modulation during solar cycle 22.}
\label{GLEFOR}
\end{figure}

The larger Forbush decreases during the time span of the ISCCP data (1984-2005) 
have been examined to see if they could be correlated with changes in the LCC. 
Most of these give relatively small changes in the CR intensity compared to 
the 11 year solar cycle modulation. Similar changes in the rates 
in the Nagoya muon detector \cite{Nagoya} were observed to those in a 
neutron monitor at the same VRCO.        
The globally averaged cloud cover change was taken as the difference between 
the LCC, using the ISCCP D2 data, in the month of the decrease and the average 
of the 3 preceding months. For some large shorter duration events the D1 data 
were used, taking the difference between the average LCC during 14 days 
before the event and seven days after.   
Figure \ref{GLEFOR} shows the change in the LCC anomaly for each Forbush decrease 
plotted against the change in the Oulu neutron monitor count rate averaged 
over the duration of the decrease. The data below an Oulu count rate change of $9\%$, 
which is roughly half the solar modulation during solar cycle 22, are too 
statistically imprecise to be conclusive. The statistical errors were determined 
from the RMS deviation of these points about the mean. However, the four points above 
a  counting rate change of $9\%$ have a mean LCC anomaly change of $0.68\pm0.45\%$. 
This is compatible with the dashed line showing no correlation between the LCC and 
CR rate changes but it is 2.8 standard deviations above the value of $-0.6\%$ expected 
had there been a correlation similar to that seen in figure \ref{figsec}.    

A further attempt was made to correlate monthly fluctuations in the neutron monitor rates
with those in the LCC. For each of the LCC and Climax neutron monitor monthly averages  
a linear extrapolation from 7 of the measurements was made to the eighth. The 
fluctuation was then taken to be the difference between the eighth measurement and 
the extrapolated value. The regression line fitted to the plot of fluctuations in the 
LCC against the fluctuations in the Climax data had the form 
$\Delta \rm{LCC}=-0.0098\pm0.019 \Delta N_C/N_C$, with correlation coefficient -0.03  
indicating a poor correlation. Here $\Delta \rm{N_C/N_C}$ in per cent is the fluctuation 
in the Climax count rate.   If the dip in LCC shown in figure \ref{fig1} is due to ionization 
from cosmic rays as hypothesised in \cite{PBB,MS1}, the curve fitted to the data in 
figure \ref{figsec} would predict that this line should have the form
$\Delta \rm{LCC}=-0.048 \Delta \rm{N_C/N_C}$ i.e. a slope which is 2 standard 
deviations greater than that obtained from these fluctuations.  

In conclusion, it is statistically improbable that the Forbush decreases 
are compatible with the hypothesis of a correlation between LCC and ionization 
as proposed in \cite{PBB,MS1}. Hence Forbush decreases do not provide 
evidence which can be used to corroborate 
such a hypothesis.  There have been previous reports of observations of correlations 
between cloud cover and Forbush decreases \cite{Harrison,Vere}. These seem to 
be incompatible with our observations although the statistical precision of the 
data is not powerful. 


\section{Conclusions}
The dip in amplitude of $1.28\%$ in the low altitude cloud cover
noted in references \cite{PBB,MS1} in solar cycle 22 (peaking in 1990)
has also been seen in this analysis. This dip anti-correlates in
shape with the observed mean daily sun spot number i.e. correlates
with the change in cosmic ray intensity due to solar modulation. 
The dip is less evident in the following solar cycle 23 although it is 
possibly present in the tropical regions of the Earth. If the 
correlation noted in \cite{PBB,MS1} and its hypothesised causal 
connection between low cloud cover and ionization are real, it is shown 
that the magnitude of the effect implies that a large 
fraction of the low cloud cover is formed by ionization. However, no 
evidence could be found of changes in the cloud cover from  
known changes in the cosmic ray ionization rate.      

In conclusion, no corroboration of the claim of a causal connection 
between the changes in ionization and cloud cover, made in \cite{PBB,MS1}, 
could be found in this investigation. From the distribution of the depth 
of the dip in solar cycle 22 with geomagnetic latitude (the VRCO) 
we find that, averaged over the whole Earth, less than 23$\%$
of the dip comes from the solar modulation of the 
cosmic ray intensity, at the 95$\%$ confidence level. 
This implies that, if the dip represents a real correlation, more 
than $77\%$ of it is caused by a source other than ionization   
and this source must be correlated with solar activity.

\section{Acknowledgment}
We wish to thank the WDC for Cosmic Rays (Solar-Terrestrial Environment
Laboratory, Nagoya University) for their compilation of the Cosmic-Ray Neutron 
Data (CAWSESDB-J-OB0061) \cite{Watanabe}.  This database was constructed as a 
part of CAWSES Space Weather International Collaborative Research Database 
in Japan. We also thank the Cosmic Ray section for the provision of the data 
from the Nagoya Muon Telescope \cite{Nagoya}. In addition we thank the 
ISCCP \cite{ISCCP} for the provision of the cloud data. We are grateful 
to the referees of the paper for their thoughtful comments which helped 
us to improve it considerably. We also thank A. Erlykin and R.G. Harrison 
for helpful discussions.





\end{document}